\documentclass{eptcs}
\providecommand{\event}{GandALF 2010} 
\usepackage{breakurl}             

\usepackage{tikz}
\usetikzlibrary{arrows,automata}
\usepackage{amsmath}
\usepackage{amssymb}
\usepackage{amsthm}

\newtheorem{theorem}{Theorem}
\newtheorem{lemma}[theorem]{Lemma}
\newtheorem{definition}[theorem]{Definition}
\newtheorem{corollary}[theorem]{Corollary}
\newtheorem{remark}[theorem]{Remark}

\title{Formats of Winning Strategies for Six Types of \\ Pushdown Games}
\author{Wladimir Fridman
\institute{Chair of Computer Science 7\\
RWTH Aachen University\\
Aachen, Germany}
\email{fridman@automata.rwth-aachen.de}
}
\def\titlerunning{Formats of Winning Strategies for Six Types of Pushdown Games}
\def\authorrunning{W. Fridman}
\begin{document}
\maketitle

\begin{abstract}
The solution of parity games over pushdown graphs (Walukiewicz '96) was the first step towards an effective theory of infinite-state games. It was shown that winning strategies for pushdown games can be implemented again as pushdown automata. We continue this study and investigate the connection between game presentations and winning strategies in altogether six cases of game arenas, among them realtime pushdown systems, visibly pushdown systems, and counter systems. In four cases we show by a uniform proof method that we obtain strategies implementable by the same type of pushdown machine as given in the game arena. We prove that for the two remaining cases this correspondence fails. In the conclusion we address the question of an abstract criterion that explains the results.
\end{abstract}


\section{Introduction}

When we look at nonterminating reactive systems, two agents, a \textit{controller} and the \textit{environment}, can be identified interacting with each other. At each point in time $i$, the \textit{environment} executes an action $\alpha(i)\in\Sigma_{\!E}$ which is directly consumed by the \textit{controller} and responded by an action $\beta(i)\in\Sigma_{\!C}$, for $\Sigma_{\!E}$ and $\Sigma_{\!C}$ being finite sets of actions that can be chosen by \textit{environment} and \textit{controller} respectively. A \textit{system behavior} can be described by two infinite sequences $\alpha=\alpha(0)\alpha(1)...$ and $\beta=\beta(0)\beta(1)...$ produced by the two agents, thus a system behavior is an infinite sequence $\alpha \choose \beta$  of pairs $\alpha(i) \choose \beta(i)$. A \textit{system specification} is a language $L\subseteq(\Sigma_{\!E}\times \Sigma_{\!C})^\omega$ consisting of all correct system behaviors.

\textit{Church's Problem}, first stated by A. Church \cite{Church57,Church63}, is to synthesize a finite state controller from a given regular system specification. So, when given a regular specification language $L$ the question to be answered is if there is an automaton (transducer) that transforms every input $\alpha\in\Sigma_{\!E}^\omega$ letter by letter into an output $\beta\in\Sigma_{\!C}^{\omega}$, such that the specification is fulfilled, i.e., ${\alpha\choose\beta} \in L$, and if the answer is positive, such an automaton should (automatically) be constructed. 

Church's Problem can be formulated in the framework of \textit{infinite two-player games} as a slightly modified version of a \textit{Gale-Stewart game} \cite{GaSt1953}.
The winning condition is given by an $\omega$-language $L$. The winner of a play is established by testing it for membership in $L$, \textit{Player 0} (controller) wins a play if it is contained in $L$ and \textit{Player 1} (environment) wins if it is not. 

A \textit{strategy} for a player is a function mapping a finite play prefix to a letter the player should choose next, $f_1\colon \bigcup_{n\in\mathbb{N}}(\Sigma_{\!E}\times\Sigma_{\!C})^{n} \rightarrow \Sigma_{\!E}$ for Player 1 and $f_0\colon \bigcup_{n\in\mathbb{N}}(\Sigma_{\!E}\times\Sigma_{\!C})^{n}(\Sigma_{\!E}\times\{\star\}) \rightarrow \Sigma_{\!C}$ for Player 0 where the symbol $\star\notin\Sigma_{\!C}$ serves as placeholder. A strategy $f$ is \textit{winning} for a player if it guarantees that the player wins any play if he always acts according to $f$. To solve Church's Problem one has to find the winner and a winning strategy.

The first solution was offered by B{\"u}chi and Landweber \cite{BuchiLandweber69}, who established the following fundamental result on regular games.

\begin{theorem}[B{\"u}chi, Landweber 1969]
 For each MSO-definable game either Player 0 or Player 1 has a finite-state winning strategy and the winner and a finite-state machine realizing a winning strategy can be computed.
\end{theorem}

\noindent This result was refined in two papers where a close conceptual connection between the formats of winning conditions and winning strategies has been established. It was shown for several regular subclasses that specifications and winning strategies can be defined in corresponding formats. For $\mathcal{L}$ being one of the logics MSO, FO($<$), FO(S), FO($<$)+MOD or strictly bounded logic, it holds that each $\mathcal{L}$-definable game is determined with an $\mathcal{L}$-definable winning strategy \cite{DBLP:conf/csl/RabinovichT07}. Each game defined by an aperiodic $\omega$-language is determined by a winning strategy which can again be realized by an aperiodic transducer \cite{DBLP:conf/dlt/Selivanov07}.

In this paper we pursue this study and consider the connection between winning conditions and winning strategies for context-free games. We shall show that games defined by the following types of pushdown machines are determined with winning strategies realizable by the same types of pushdown machine: (1) deterministic; (2) deterministic visibly; (3) deterministic realtime; (4) deterministic one-counter. This statement is shown by a refinement of an automata-theoretic approach due to Kupferman and Vardi \cite{DBLP:conf/cav/KupfermanV00}. On the other hand we indicate two cases where this statement fails, namely blind one-counter and visibly one counter games.

This paper is structured as follows: in the subsequent section \ref{Prelim} we introduce the considered types of pushdown games and state our main result. Section \ref{SolvingPDG} recalls the technique of \cite{DBLP:conf/cav/KupfermanV00} which is adapted to prove the theorems in section \ref{Proof}. We conclude by a brief outline of our efforts concerning a generalization of the result.


\section{Preliminaries and Main Result}\label{Prelim}

For any set $X$ the power set will be denoted by $\mathcal{P}(X)$, $\mathbb{N}$ denotes the set of non-negative integers. For an alphabet $\Sigma$, $\Sigma^{\ast}$ denotes the set of finite words over $\Sigma$ and $\Sigma^{\omega}$ the set of infinite words over $\Sigma$. The empty word is denoted by $\varepsilon$. For a word $w\in\Sigma^\ast$ the reverse of $w$ is denoted by $w^R$. For $\alpha\in \Sigma^{\ast}\cup\Sigma^{\omega}$ and $n\in \mathbb{N}$ we write $\alpha(n)$ for the $n$-th letter  of $\alpha$. For an integer $k>0$ let $[k]$ denote the set $\{0,...,k-1\}$.

\subsubsection*{Pushdown Games, Pushdown Strategies}

\begin{definition} A \textit{pushdown machine} (PDM) is a tuple $\mathcal{M} = (Q, \Sigma, \Gamma, \delta, q_{in}, \bot)$ 
where $Q$ is a finite set of states, $\Sigma$ is a finite input alphabet,  $\Gamma$ is a finite pushdown alphabet, $\bot\notin\Gamma$ is the initial pushdown symbol (let $\Gamma_{\!\bot}=\Gamma\cup\{\bot\}$), $q_{in}\in Q$ is the initial state and $\delta$ is a mapping from
$Q\times(\Sigma\cup\{\varepsilon\})\times\Gamma_{\!\bot}$ into \linebreak $\mathcal{P}(Q\times\Gamma_{\!\bot}^{*})$.
A PDM is \textit{deterministic} (DPDM), if $\forall q \in Q, \forall a \in \Sigma, \forall A \in \Gamma_{\!\bot}$: 
$|\delta(q,a,A)|+|\delta(q,\varepsilon,A)|\leq 1$.
\end{definition}

\noindent The initial pushdown symbol $\bot$ can neither be written on the stack nor be deleted from the stack. A stack content is a word from $\Gamma^{\ast}\bot$, we assume the leftmost symbol to be the top of the stack. 

A \textit{configuration} is a pair $(q, \gamma)$ consisting of a state $q\in Q$ and a stack content $\gamma\in\Gamma^{\ast}\bot$. The stack height of a configuration $(q, \gamma)$ is defined as $sh((q, \gamma))=|\gamma|$. We write $(q,A\gamma) \overset{a}{\mapstochar\relbar} (q', \gamma'\gamma)$, if $(q',\gamma')\in\delta(q,a,A)$ for $a\in\Sigma\cup\{\varepsilon\}$, $\gamma,\gamma'\in\Gamma_{\!\bot}^{\ast}$ and $A \in \Gamma_{\!\bot}$.

For a finite word $w=w(0)...w(n)\in \Sigma^{\ast}$, a finite sequence $\rho=(q_0,\gamma_0 )...(q_m,\gamma_m)$ of configurations is a (\textit{finite}) \textit{run} of a PDM $\mathcal{M}$ on $w$ iff (1) $(q_0, \gamma_0)=(q_{in}, \bot)$ and (2) for all $0\leq i < m$ exists $a_i \in \Sigma\cup\{\varepsilon\}$, such that $(q_i,\gamma_i)\overset{a_i}{\mapstochar\relbar}(q_{i+1},\gamma_{i+1})$ and $a_0...a_m = w$. 
For an $\omega$-word $\alpha=\alpha(0)\alpha(1)... \in \Sigma^{\omega}$ an infinite sequence of configurations $\rho= (q_0,\gamma_0 )(q_1,\gamma_1)...$ is an (\textit{infinite}) \textit{run} of $\mathcal{M}$ on $\alpha$ iff (1) $(q_0, \gamma_0)=(q_{in}, \bot)$ and (2) for all $i \in \mathbb{N}$ exists $a_i \in \Sigma\cup\{\varepsilon\}$, such that $(q_i,\gamma_i)\overset{a_i}{\mapstochar\relbar}(q_{i+1},\gamma_{i+1})$ and $a_0a_1... = \alpha$.

A \textit{pushdown graph} of a PDM $\mathcal{M}$ is a graph $G(\mathcal{M})=(V_{\mathcal{M}}, E_{\mathcal{M}})$ where $V_{\mathcal{M}}=\{(q,\gamma) \ | \ q \in Q$, $ \gamma \in \Gamma^{\ast}\bot\}$ and $E_{\mathcal{M}}\subseteq V_{\mathcal{M}} \times (\Sigma\cup\{\varepsilon\}) \times V_{\mathcal{M}}$, $((q,\gamma),a,(q',\gamma')) \in E_{\mathcal{M}}$ if $(q,\gamma)\overset{a}{\mapstochar\relbar}(q',\gamma')$, for $a\in\Sigma\cup\{\varepsilon\}.$

\begin{definition} A \textit{pushdown automaton} (PDA) is a tuple $\mathcal{A} = (\mathcal{M}^\mathcal{A}, F)$ where $\mathcal{M}^\mathcal{A}$ is a PDM and $F\subseteq Q$ is a set of \textit{final states}. The (finitary) language recognized by $\mathcal{A}$ is $L(\mathcal{A})=\{w\in\Sigma^{\ast}|$ there exists a run $\rho=(q_0,\gamma_0 )...(q_m,\gamma_m)$ of $\mathcal{A}$ on $w$, such that $q_m\in F\}$.
\end{definition}

The class of context-free (finitary) languages, denoted by CFL, is exactly the class of languages accepted by pushdown automata.

We now define pushdown automata accepting $\omega$-words which were first introduced in \cite{DBLP:journals/jcss/CohenG77, DBLP:journals/jcss/CohenG77a}. For an infinite run $\rho$ of a PDM let Inf($\rho$) denote the set of states seen infinitely often in $\rho$, i.e., Inf$(\rho)=\linebreak\{q \in Q \ | \ \forall i \in \mathbb{N} \ \exists j>i, \gamma_j \in \Gamma_{\!\bot}^{\ast}: \rho(j)=(q,\gamma_j)\}$, and $Steps_{\rho}=\{n\in\mathbb{N} \ | \ \forall m\geq n: sh(\rho(m))\geq sh(\rho(n))\}$. Note that $Steps_{\rho}$ is infinite for every infinite run $\rho$. For a set $Steps=\{n_i \ | \ i\in \mathbb{N}\}\subseteq \mathbb{N}$ with $n_0<n_1<n_2<...$ and an $\omega$-word $\rho$ over any alphabet, let $\rho|_{Steps}=\rho(n_0)\rho(n_1)\rho(n_2)...$ .

Let $col\colon Q \rightarrow [k]$ be a \textit{priority function} assigning to each state of a PDM $\mathcal{M}$ a natural number. We consider two kinds of accepting conditions for $\omega$-pushdown automata. A run $\rho$ satisfies the \textit{parity condition} if the minimal priority seen infinitely often in $\rho$ is even. A run $\rho$ satisfies the \textit{stair parity condition} \cite{DBLP:conf/fsttcs/LodingMS04} if the minimal priority seen infinitely often in the subsequence $\rho|_{{Steps}_\rho}$ is even.

\begin{definition} An \textit{$\omega$-pushdown automaton} ($\omega$-PDA) is a tuple $\mathcal{A} = (\mathcal{M}^\mathcal{A}, col)$ where $\mathcal{M}^\mathcal{A}$ is a PDM and $col\colon Q \rightarrow [k]$ is a priority function. 
A \textit{parity pushdown automaton} (parity-PDA) accepts an $\omega$-word $\alpha\in\Sigma^{\omega}$ if there exists a run $\rho$ of $\mathcal{A}$ on $\alpha$, such that min$\{col(q) \ | \ q\in$ Inf$(\rho)\}$ is even.
A \textit{stair parity pushdown automaton} (parity-StPDA) accepts an $\omega$-word $\alpha\in\Sigma^{\omega}$ if there exists a run $\rho$ of $\mathcal{A}$ on $\alpha$, such that min$\{col(q) \ | \ q\in$ Inf$(\rho|_{{Steps}_\rho})\}$ is even.
\end{definition}

The class of $\omega$-languages accepted by parity pushdown automata is the class of $\omega$-context-free languages, denoted by CFL$_\omega$.

For a PDM $\mathcal{M}$ consider a partition $Q_1\cup Q_0$ of the set of states $Q$. It induces a partition $V_\mathcal{M}=V_1\cup V_0$ where $V_1=\{(q,\gamma)\in V_\mathcal{M} \ | \ q\in Q_1\}$ and $V_0=\{(q,\gamma)\in V_\mathcal{M} \ | \ q\in Q_0\}$. A \textit{pushdown game graph} is defined for a PDM $\mathcal{M}$ with a partition $Q=Q_1\cup Q_0$ as $G(\mathcal{M})=(V_1\cup V_0, E_{\mathcal{M}})$. An $\omega$-PDA $\mathcal{A}$ with a partition $Q=Q_1\cup Q_0$ induces a \textit{pushdown game} $\mathcal{G}(\mathcal{A})=(G(\mathcal{M}^\mathcal{A}),col)$ (parity game or stair parity game respectively played on a pushdown game graph) where Player $i$ chooses a transition if the current configuration $(q,\gamma)$ is in $V_i$ for $i\in\{0,1\}$. The initial configuration of $\mathcal{G}(\mathcal{A})$ is $(q_{in},\bot)$. Player 0 wins a play $\rho\in(Q\times \Gamma^{\ast}\bot)^\omega$ starting in the initial configuration if $\rho$ satisfies the parity condition or the stair parity condition respectively, otherwise Player 1 wins. Note that Church's Problem with a specification $L$ given by an $\omega$-PDA introduces a pushdown game.

We define a \textit{pushdown strategy} $\mathcal{S}$ as a deterministic PDA with output, $\mathcal{S}=(Q, \Sigma_{i}, \Sigma_o, \Gamma, \delta, q_{in}, \bot)$ where $\mathcal{M}^\mathcal{S}=(Q, \Sigma_{i}, \Gamma, \delta, q_{in}, \bot)$  is a DPDM, $\Sigma_o$ is a finite output alphabet and the transition function $\delta$ is extended such that it is a mapping from $Q \times (\Sigma_i \cup \{\varepsilon\}) \times \Gamma_{\!\bot}$ into $Q \times {\Gamma_{\!\bot}}^{\ast} \times (\Sigma_o \cup \{\varepsilon\})$. If $\delta(q,a,A)=(q',\gamma, x)$ then the automaton being in state $q$ with $A$ on the top of the stack proceeds via an input symbol $a$ to state $q'$ changing the top of the stack to $\gamma$ and outputs $x$. If $a=\varepsilon$ then $\mathcal{S}$ performs an $\varepsilon$-transition and if $x=\varepsilon$ then the automaton outputs nothing. Another possibility to define a strategy realized in terms of pushdown machines is to specify a set $\{\mathcal{A}_x \ | \ x\in\Sigma_o\}$ of DPDA where the languages $L(\mathcal{A}_x)$ are pairwise disjoint and for every $x\in\Sigma_o$, $\mathcal{A}_x$ accepts all the finite play prefixes where the next choice should be $x$.

\subsubsection*{Types of Context-Free ($\omega$-)Languages and Main Result}

There are various classes of context-free and $\omega$-context-free languages conceivable which can be described by a set of properties of the underlying pushdown machines defining those classes.

\textit{Determinism}. We denote the class of deterministic context-free (finitary) languages by DCFL. The class of deterministic $\omega$-context-free languages accepted by deterministic parity-PDA is denoted by DCFL$_\omega$. The class of languages accepted by deterministic parity-StPDA is denoted by StDCFL$_\omega$.

\textit{Visibility}. Let $\Sigma=\Sigma_c\cup\Sigma_r\cup\Sigma_{int}$ be an alphabet partitioned into three disjoint alphabets. $\Sigma_c$ is a set of \textit{calls}, $\Sigma_r$ a set of \textit{returns}, $\Sigma_{int}$ is a set of \textit{internal actions}. We denote the tuple $\langle\Sigma_c$, $\Sigma_r$, $\Sigma_{int}\rangle=\widetilde{\Sigma}$ a \textit{visibly pushdown alphabet}. A \textit{visibly pushdown machine} (VPM) is a PDM $\mathcal{M} = (Q, \widetilde{\Sigma}, \Gamma, \delta, q_{in}, \bot)$ where $\widetilde{\Sigma}$ is a visibly pushdown alphabet and the transition function is composed of three functions $\delta=\delta_c\cup\delta_r\cup\delta_{int}$ where $\delta_c\colon Q\times\Sigma_c \rightarrow \mathcal{P}(Q\times\Gamma)$,  $\delta_r\colon Q\times\Sigma_r\times\Gamma_{\!\bot}\rightarrow \mathcal{P}(Q)$ and $\delta_{int}\colon Q\times\Sigma_{int}\rightarrow \mathcal{P}(Q)$.
VPL (DVPL) denotes the class of (deterministic) visibly pushdown finitary languages accepted by PDA (DPDA) where the corresponding PDM is a VPM. Classes of visibly pushdown $\omega$-languages are denoted by VPL$_\omega$, StVPL$_\omega$ and for the deterministic case DVPL$_\omega$, StDVPL$_\omega$. 

\begin{remark}[\cite{DBLP:conf/stoc/AlurM04}]
 For finitary visibly pushdown languages the classes are equivalent DVPL $=$ VPL.
 For visibly pushdown $\omega$-languages the inclusion is strict DVPL$_\omega$ $\subsetneq$ VPL$_\omega$.
\end{remark} 

\begin{remark}[\cite{DBLP:conf/fsttcs/LodingMS04}]\label{VPLclasses}
 VPL$_\omega=$ StVPL$_\omega=$ StDVPL$_\omega$.
\end{remark} 

\begin{lemma}\label{DPDAstDPDA}
 For every parity-DPDA $\mathcal{A}$ an equivalent parity-StDPDA $\mathcal{A}'$ can be constructed, such that $L(\mathcal{A})=L(\mathcal{A}')$.
\end{lemma}

\begin{remark}
$\ $
\begin{enumerate}
 \item VPL$_\omega$ $\nsubseteq$ DCFL$_\omega$, DCFL$_\omega$ $\nsubseteq$ VPL$_\omega$ and  VPL$_\omega$ $\cap$ DCFL$_\omega$ $\neq \emptyset$
 \item VPL$_\omega$ $\subseteq$ StDCFL$_\omega$
 \item DCFL$_\omega$ $\subseteq$ StDCFL$_\omega$
 \item VPL$_\omega$ $\cup$ DCFL$_\omega$ $\neq$ StDCFL$_\omega$
\end{enumerate}
\end{remark} 

\begin{proof}
 (1.) Consider $L=\{a^nba^nb^\omega \ | \ n\in \mathbb{N}\}\subseteq\{a,b\}^\omega$. Obviously $L\in$ DCFL$_\omega$, but for any partition of $\{a,b\}$ in calls, returns and internal actions $L\notin$ VPL$_\omega$. On the other hand the class VPL$_\omega$ is contained in $\mathcal{B}(\Sigma_3)$ (Boolean closure of the third level of the Borel hierarchy) \cite{DBLP:conf/fsttcs/LodingMS04} which exceeds $\mathcal{B}(\Sigma_2)$ wherein DCFL$_\omega$ is contained \cite{DBLP:journals/jcss/CohenG77, DBLP:journals/jcss/CohenG77a}. Moreover, every deterministic visibly pushdown $\omega$-language is also in DCFL$_\omega$.
 (2.) With remark \ref{VPLclasses} and the fact that every VPM is also a PDM it holds that VPL$_\omega=$ StDVPL$_\omega\subseteq$ StDCFL$_\omega$.
 (3.) Follows from lemma \ref{DPDAstDPDA}.
 (4.) Let $L_1\subseteq\{c,r\}^\omega$ with $L_1\in$ VPL$_\omega\setminus$DCFL$_\omega$ and $L_2=\{a^nba^n\alpha \ | \ \alpha\in L_1\}\subseteq\{a,b,c,r\}^\omega$. Obviously $L_2\notin$ DCFL$_\omega$ and $L_2\notin$ VPL$_\omega$, but it is easy to verify that $L_2\in$ StDCFL$_\omega$.
\end{proof}

\textit{Realtime}. A DPDM $\mathcal{M}$ is called \textit{realtime} if the corresponding transition function $\delta^\mathcal{M}$ is a mapping from $Q\times\Sigma\times\Gamma_{\!\bot}$ into $\mathcal{P}(Q\times\Gamma_{\!\bot}^{*})$, i.e., if $\delta^\mathcal{M}$ contains no $\varepsilon$-transitions.
The corresponding classes of languages are denoted by realtime-DCFL, realtime-DCFL$_\omega$ and realtime-StDCFL$_\omega$. Note that every VPM is realtime.

\textit{Counter}. A DPDM $\mathcal{M}$ is a deterministic \textit{one-counter machine} (D1CM) if the stack alphabet contains only one symbol, $|\Gamma^\mathcal{M}|=1$. We denote the classes of one-counter languages by D1CL, D1CL$_\omega$, StD1CL$_\omega$.

\textit{Blindness}. A D1CM $\mathcal{M}$ is called \textit{blind} (DB1CM) if $\forall q, q' \in Q^\mathcal{M}, \forall a \in \Sigma\cup\{\varepsilon\}$: if $\delta^\mathcal{M}(q,a,\bot)=(q',A^n\bot)$ for some $n\geq0$, then $\delta^\mathcal{M}(q,a,A)=(q',A^nA)$, i.e., every transition which is enabled with empty stack is also enabled with the stack being nonempty. Thus, a blind one-counter cannot check if its stack is empty or not.
The classes of blind one-counter languages are denoted by DB1CL, DB1CL$_\omega$ and StDB1CL$_\omega$.

A format of a PDM can be regarded as a combination of such properties defining a class of context-free languages. For example we can define the class of deterministic visibly one-counter languages denoted by DV1CL.

\begin{theorem}
$\ $ 
\begin{enumerate}
 \item {DCFL$_\omega$}-games and {StDCFL$_\omega$}-games are determined with {DCFL} winning strategies.
 \item {DVPL$_\omega$}-games and {StDVPL$_\omega$}-games are determined with {DVPL} winning strategies.
 \item {realtime-DCFL$_\omega$}-games and  {realtime-StDCFL$_\omega$}-games are determined with {realtime-DCFL} winning strategies.
 \item {D1CL$_\omega$}-games and {StD1CL$_\omega$}-games are determined with {D1CL} winning strategies.
\end{enumerate}
\end{theorem}

\begin{theorem}
$\ $ 
\begin{enumerate}
 \item {DB1CL$_\omega$}-games and {StDB1CL$_\omega$}-games are determined, however {DB1CL} winning strategies do not suffice.
 \item {DV1CL$_\omega$}-games and {StDV1CL$_\omega$}-games are determined, however {DV1CL} winning strategies do not suffice.
\end{enumerate}
\end{theorem}
 
\subsubsection*{Alternating Two-Way Tree Automata}

For a given set $\Gamma$ of directions a $\Gamma$\textit{-tree} is a prefix closed set $T\subseteq\Gamma^{\ast}$, i.e., for $\gamma\in \Gamma^{\ast}$ and $A\in\Gamma$, if $\gamma A\in T$, then also $\gamma\in T$ ($\gamma A$ is called a \textit{child} of $\gamma$ and $\gamma$ is the \textit{parent} of $\gamma A$). The elements of $T$ are called \textit{nodes} and the empty word $\varepsilon$ is the \textit{root} of $T$. If $T=\Gamma^{\ast}$, it is called a \textit{full infinite tree}. A \textit{labeled} $\Gamma$-tree over an alphabet $\Sigma$ is a pair $(T,\lambda)$ where $T$ is a $\Gamma$-tree and $\lambda$ is a mapping from $T$ into $\Sigma$ assigning to each node a symbol from $\Sigma$.

For a finite set $X$, let $\mathcal{B}^{+}(X)$ denote the set of positive Boolean formulas over $X$ where the formulas \textit{true} and \textit{false} are also allowed. A set $Y\subseteq X$ \textit{satisfies} a formula $\theta\in\mathcal{B}^{+}(X)$ iff $\theta$ is true when assigned \textit{true} to all elements in $Y$ and \textit{false} to all elements in $X\setminus Y$.

An \textit{alternating two-way parity tree automaton} (A2TA) $\mathcal{A}$ over $\Sigma$-labeled $\Gamma$-trees is a tuple \linebreak $(Q, \Sigma, q_{in}, \delta, col)$ where $Q$ is a finite set of states, $\Sigma$ is a finite input alphabet, $q_{in}\in Q$ is the initial state, \linebreak $col\colon Q\rightarrow[k]$ is  a priority function and $\delta$ is a mapping from $Q\times \Sigma$ into 
$\mathcal{B}^{+}(\{\uparrow,\downarrow_A,N \ | \ A\in\Gamma\}\times Q)$ where the set $Dir=\{\uparrow,\downarrow_A,N \ | \ A\in\Gamma\}$ serves for navigation through the tree. For all $\gamma\in \Gamma^{\ast}$, $A\in \Gamma$, we define $\gamma.N = \gamma$, $\gamma.$\!$\downarrow_A=\gamma A$, $\gamma A.$\!$\uparrow = \gamma$. A run of $\mathcal{A}$ on a $\Sigma$-labeled $\Gamma$-tree $(T,\lambda)$ is a $(Q\times T)$-labeled \linebreak $\Xi$-tree $(T_r,\lambda_r)$, for some set of directions $\Xi$,  where the following conditions are fulfilled:\linebreak (1) $\varepsilon\in T_r$ and $\lambda_r(\varepsilon)=(q_{in},\varepsilon)$, (2) let $\xi\in T_r$ with $\lambda_r(\xi)=(q,\gamma)$ and $\delta(q,\lambda(\gamma))=\theta$, then there is a set $\{(dir_1, q_1), (dir_2, q_2), ..., (dir_n, q_n)\}\subseteq \{Dir\times Q\}$ that satisfies $\theta$ and for all $1\leq i\leq n$ there is $x_i\in\Xi$ such that $\xi x_i \in T_r$ and $\lambda_r(\xi x_i)=(q_i,\gamma.dir_i)$. A run $(T_r, \lambda_r)$ is \textit{accepting} iff all its infinite paths $\rho\in(Q\times T)^\omega$ satisfy the parity condition.


\section{Solving Pushdown Games}\label{SolvingPDG}

In this section we recall some results on pushdown games, in particular the technique proposed by Kupferman and Vardi \cite{DBLP:conf/cav/KupfermanV00} that comprises a reduction to the emptiness problem for alternating two-way parity tree automata, which can be applied to solve deterministic pushdown games. First note that in general pushdown games cannot be solved.

\begin{remark}[Finkel \cite{DBLP:journals/tcs/Finkel01a}]
For nondeterministic context-free languages $L \in$ CFL$_\omega$ it is undecidable to determine which player has a winning strategy in the Gale-Stewart game defined by $L$. 
\end{remark}

\noindent The proof of this fact uses the undecidability of the universality problem for context-free languages. From this we can directly conclude that Church's Problem for the class CFL$_\omega$ is undecidable. On the other hand Walukiewicz showed that deterministic pushdown games can be solved, by a method reducing pushdown games to parity games on finite game graphs for which the determinacy and feasibility to construct the winning regions and the memoryless winning strategies effectively are known \cite{DBLP:conf/focs/EmersonJ91}. 

\begin{theorem}[Walukiewicz \cite{DBLP:conf/cav/Walukiewicz96}]
Deterministic parity pushdown games are determined with deterministic pushdown winning strategies.
\end{theorem}

The idea of \cite{DBLP:conf/cav/KupfermanV00} is to simulate a pushdown game on a full $\Gamma_{\!\bot}$-labeled $\Gamma$-tree (representing all possible stack contents of the corresponding DPDM) by the use of alternating two-way parity tree automata. Pushdown winning strategies can be derived from the A2TA simulating the pushdown game  by testing it for nonemptiness. The essential step thereby is the translation of an A2TA into an equivalent nondeterministic one-way parity tree automaton (N1TA) \cite{DBLP:conf/icalp/Vardi98}.

Let us recall the construction which can similar be found  in \cite{DBLP:conf/dagstuhl/Cachat01}. 

\paragraph*{A2TA simulating a pushdown game.}
Let $\mathcal{A}=(Q^\mathcal{A}, \Sigma, \Gamma, \delta^\mathcal{A}, q_{in}^\mathcal{A}, \bot, col^\mathcal{A})$ be a parity-DPDA with a partition $Q^\mathcal{A}=Q_0\cup Q_1$ defining a pushdown game $\mathcal{G}(\mathcal{A})=(G(\mathcal{M}^\mathcal{A}),col^\mathcal{A})$. $\mathcal{A}$ can be assumed in the following normal form where all push-transitions are of the form $\delta^\mathcal{A}(q,a,A)=(q',A'A)$, i.e., at most one stack symbol can be pushed on the stack in a transition step, furthermore skip-transitions are of the form $\delta^\mathcal{A}(q,a,A)=(q',A)$ and pop-transitions $\delta^\mathcal{A}(q,a,A)=(q',\varepsilon)$.

From $\mathcal{A}$ we define an A2TA $\mathcal{B}$ which simulates the pushdown game $\mathcal{G}(\mathcal{A})$. For all $q\in Q^{\mathcal{A}}$, $A\in\Gamma_{\!\bot}$, let $\delta$ be the following function:
$$ \delta(q,A)=\begin{cases}
                                     {\displaystyle \bigvee_{\substack{\delta^\mathcal{A}(q,a,A)\\=(q',A'A)}}{(\downarrow_{A'},q')} \vee 
				\bigvee_{\substack{\delta^\mathcal{A}(q,a,A)\\=(q',A)}}{(N,q')} \vee
				\bigvee_{\substack{\delta^\mathcal{A}(q,a,A)\\=(q',\varepsilon)}}{(\uparrow,q')}},  & \text{if }q\in Q_0\\
				& \\
				{\displaystyle \bigwedge_{\substack{\delta^\mathcal{A}(q,a,A)\\=(q', A'A)}}{(\downarrow_{A'},q')} \wedge 
				\bigwedge_{\substack{\delta^\mathcal{A}(q,a,A)\\=(q', A)}}{(N,q')} \wedge
				\bigwedge_{\substack{\delta^\mathcal{A}(q,a,A)\\=(q', \varepsilon)}}{(\uparrow, q')}}, & \text{if }q\in Q_1.
                                      \end{cases}$$
Then $\mathcal{B}=(Q^\mathcal{B}, \Gamma_{\!\bot}, q_{in}^\mathcal{B}, \delta^\mathcal{B}, col^\mathcal{B})$ is defined as follows:

\begin{itemize}
 \item $Q^\mathcal{B}=Q^\mathcal{A}\cup\overline{Q}\cup\{q_{in}^\mathcal{B}\}$ where $\overline{Q}=\{\overline{q}_{\!A} \ | \ A\in \Gamma\}$, $q_{in}^\mathcal{B}\notin Q^\mathcal{A}\cup\overline{Q}$ and $\overline{q}_{\!A}\notin Q^\mathcal{A}$ for all $A\in \Gamma$ 

 \item the transition function $\delta^\mathcal{B}$
   \begin{itemize}
	  \item $\delta^\mathcal{B}(q,A)=\delta(q,A)$ if $q\in Q^\mathcal{A}$
	  \item $\delta^\mathcal{B}(q_{in}^\mathcal{B},\bot)=(N,q_{in}^\mathcal{A}) \wedge \bigwedge_{A\in\Gamma}{(\downarrow_{A},\overline{q}_{\!A})}$
  	\item $\delta^\mathcal{B}(\overline{q}_{\!B},B')=\begin{cases}
                                      \bigwedge_{A\in\Gamma}{(\downarrow_{A},\overline{q}_{\!A})},  & \text{if }B=B'\\
																      false, & \text{if }B\neq B'.
                                      \end{cases}$
   \end{itemize}
 \item $col^\mathcal{B}(q)=col^\mathcal{A}(q)$ if $q\in Q^\mathcal{A}$, and $col^\mathcal{B}(\overline{q}_{\!A})=col^\mathcal{B}(q_{in}^\mathcal{B})=0$.
\end{itemize}

The A2TA $\mathcal{B}$ operates on the full $\Gamma_{\!\bot}$-labeled $\Gamma$-tree $T_{stack}=(\Gamma^\ast, \lambda)$ where $\lambda(\varepsilon)=\bot$ and for all $\gamma\in\Gamma^\ast$ and $A\in\Gamma$,  $\lambda(\gamma A)= A$, i.e., every node $\gamma A$ of $T_{stack}$ corresponding to the stack content $A\gamma^R\bot$ is labeled by the top of the stack $A$, and the root $\varepsilon$ corresponding to the empty stack is labeled by $\bot$. $\mathcal{B}$ simulates the pushdown transitions of $\mathcal{A}$ by moving on $T_{stack}$ and exploiting it like a stack. Using alternation, $\mathcal{B}$ can guess the best transition for Player 0 and follow each possible transition of Player 1. To verify that the input tree is $T_{stack}$, an auxiliary computation starts at the beginning of a run passing down the states from $\overline{Q}$.

\begin{theorem}
Player 0 has a winning strategy in $\mathcal{G}(\mathcal{A})$ from the initial configuration $(q_{in}^\mathcal{A},\bot)$ iff the tree $T_{stack}$ is accepted by $\mathcal{B}$.
\end{theorem}

\noindent For the proof see \cite{DBLP:conf/dagstuhl/Cachat01}. Note that since $T_{stack}$ is the only tree which can be accepted by $\mathcal{B}$, it holds that $T_{stack}\in L(\mathcal{B})$ if and only if $L(\mathcal{B})\neq\emptyset$. In order to test $\mathcal{B}$ for emptiness it is translated into an equivalent nondeterministic one-way tree automaton.

\paragraph*{From A2TA to N1TA.}
In \cite{DBLP:conf/icalp/Vardi98} the emptiness problem for A2TA is solved by a reduction to N1TA. 
We will recall the crucial steps of the construction here without giving the proofs.

Let parity-DPDA $\mathcal{A}$ and A2TA $\mathcal{B}$ be defined as above. A \textit{strategy tree} for $\mathcal{B}$ is a mapping $\tau\colon \Gamma^\ast\rightarrow \mathcal{P}(Q^\mathcal{B}\times(\Sigma\cup\{\varepsilon\})\times Dir \times Q^\mathcal{B})$ assigning to every node of the full $\Gamma$-tree a set of transitions. Intuitively, for every node (corresponding to a stack content) the labelings of a strategy tree should contain all possible transitions of Player 1 and some choices for Player 0 which unsure him to win any play. Let $St=\mathcal{P}(Q^\mathcal{B}\times(\Sigma\cup\{\varepsilon\})\times Dir \times Q^\mathcal{B})$.

Consider the tree $T_{stack}$ with the labeling augmented by a strategy, $(\Gamma^\ast, \lambda\times\tau)$. Note that a correct strategy tree $\tau$ has to be consistent, this means that the following conditions have to be satisfied, $\forall \gamma\in\Gamma^\ast, \forall (q,a,dir,q')\in \tau(\gamma)$:

\begin{enumerate}
 \item $\{(dir_1,q_1)\ | \ (q, x, dir_1, q_1)\in\tau(\gamma)\}$ satisfies $\delta^\mathcal{B}(q,\lambda(\gamma))$, i.e., the strategy satisfies the transition function $\delta^\mathcal{B}$ at every node.
 \item $\exists dir_2\in Dir, q_2\in Q^\mathcal{B}, x\in\Sigma\cup\{\varepsilon\}$: $(q', x, dir_2, q_2)\in \tau(\gamma.dir)$ or $\emptyset$ satisfies $\delta^\mathcal{B}(q',\lambda(\gamma.dir))$, i.e., the strategy is defined for state $q'$ in the node $\gamma.dir$, thus the strategy can be followed.  
 \item  $\exists dir_3\in Dir, q_3\in Q^\mathcal{B}, x\in\Sigma\cup\{\varepsilon\}$: $(q_{in}^\mathcal{B}, x, dir_3, q_3)\in \tau(\varepsilon)$ or $\emptyset$ satisfies $\delta^\mathcal{B}(q_{in}^\mathcal{B},\lambda(\varepsilon))$, i.e., for the root and the initial state a strategy is defined.
\end{enumerate}

\noindent A deterministic one-way tree automaton $\mathcal{E}_1$ over $(\Gamma_{\!\bot}\times St)$-labeled $\Gamma$-trees can be constructed which verifies this conditions. 

In the next step it must be checked that the strategy tree $\tau$ is not only consistent but also accepting, this means the parity condition $col^\mathcal{B}$ is satisfied by all consistent infinite traces in $\tau$ (a consistent infinite trace is an infinite sequence from $(Q^\mathcal{B}\times \Gamma^\ast)^\omega$ starting in $(q_{in}^\mathcal{B},\varepsilon)$ and built up by following the transitions of $\tau$). Note that in general an infinite trace produced by a strategy is bidirectional, going up and down on the tree. In order to check if a strategy tree is accepting in the one-way manner, the traces are decomposed in downwards traces and finite detours. For this purpose an annotation is defined. An \textit{annotation} for $\mathcal{B}$ is a mapping $\eta\colon \Gamma^\ast\rightarrow \mathcal{P}(Q^\mathcal{B}\times[k]\times Q^\mathcal{B})$. Given a strategy tree $\tau$, for every node of $\tau$ the annotation should contain the information about the possible finite detours at the current node and the smallest priority seen on such a detour, i.e., $(q,m,q')\in \eta(\gamma)$ means that from the node $\gamma$ and state $q$ there is a finite detour that comes back to $\gamma$ in state $q'$ with $m$ being the smallest priority seen on this detour. Let $An=\mathcal{P}(Q^\mathcal{B}\times[k]\times Q^\mathcal{B})$.

For a strategy tree $\tau$ a correct annotation $\eta$ has to satisfy the following conditions:

\begin{enumerate}
 \item if $(q,a,N,q')\in \tau(\gamma)$, then $(q,col^\mathcal{B}(q'),q')\in \eta(\gamma)$
 \item if $(q_1,m,q_2), (q_2,m',q_3)\in\eta(\gamma)$, then $(q_1,\text{min}(m, m'),q_3)\in \eta(\gamma)$
 \item if $(q,a,\downarrow_{A},q_1)\in \tau(\gamma)$ and $(q_1,a',\uparrow,q')\in \tau(\gamma A)$,
 then $(q,\text{min}\{col^\mathcal{B}(q_1),col^\mathcal{B}(q')\},q')\in \eta(\gamma)$
 \item if $(q,a,\downarrow_{A},q_1)\in \tau(\gamma)$ and $(q_1,m,q_2)\in\eta(\gamma A)$ and $(q_2,a',\uparrow,q')\in \tau(\gamma A)$,\\
 then $(q,\text{min}\{col^\mathcal{B}(q_1),col^\mathcal{B}(q')\},q')\in \eta(\gamma)$ 
\end{enumerate}

\noindent Consider a $(\Gamma_{\!\bot}\times St\times An)$-labeled full $\Gamma$-tree $(\Gamma^\ast, \lambda\times\tau\times\eta)$. A deterministic one-way tree automaton $\mathcal{E}_2$ over $(St\times An)$-labeled $\Gamma$-trees can be constructed which verifies the correctness of the annotation $\eta$ for the strategy $\tau$.

Finally an alternating one-way tree automaton $\mathcal{E}'_3$ can be constructed and then transformed into an equivalent deterministic one-way tree automaton $\mathcal{E}_3$, which evaluates the parity condition, thus identifying those trees that represent accepting runs of $\mathcal{B}$. The idea is to use the priorities stored in the annotation to not being obliged to go into the detours. For this we use pairs from $Q^\mathcal{B}\times[k]$ as states of $\mathcal{E}'_3$ with $col^{\mathcal{E}'_3}(\langle q,i\rangle)=i$. Then the transition function is defined as

$$\delta^{\mathcal{E}'_3}(\langle q, i\rangle, (S,H))=
				\bigvee_{(q,a,\downarrow_{A},q')\in S}{(\downarrow_{A},\langle q',col^\mathcal{B}(q')\rangle)} \vee 
				\bigvee_{(q,m,q')\in H}{(N,\langle q',m\rangle)}.$$

Now a deterministic one-way tree automaton $\mathcal{E}$ over $(\Gamma_{\!\bot}\times St \times An)$-labeled $\Gamma$-trees can be defined as a product of $\mathcal{E}_1$, $\mathcal{E}_2$ and $\mathcal{E}_3$ to cope with all three tasks simultaneously, checking the consistency of a strategy, verifying the correctness of an annotation for the strategy and evaluating the parity condition. Projecting out the $St$ and $An$ components from the labels leads to a N1TA $\mathcal{E}'$ over $\Gamma_{\!\bot}$-labeled $\Gamma$-trees which nondeterministically guesses the $(St\times An)$-labels and which is equivalent to $\mathcal{B}$.

\begin{theorem}[Vardi \cite{DBLP:conf/icalp/Vardi98}]
 For every A2TA $\mathcal{B}$ there exists an equivalent N1TA $\mathcal{E}'$ such that $L(\mathcal{B})=L(\mathcal{E}')$.
\end{theorem}

\noindent This result can now be applied to determine the winner in the pushdown game $\mathcal{G}(\mathcal{A})$.

\begin{corollary}
 Player 0 has a winning strategy in $\mathcal{G}(\mathcal{A})$ from the initial configuration $(q_{in}^\mathcal{A},\bot)$ 
$\Leftrightarrow$ $T_{stack}\in L(\mathcal{B})$
$\Leftrightarrow$ $L(\mathcal{B})\neq\emptyset$
$\Leftrightarrow$ $L(\mathcal{E}')\neq\emptyset$
$\Leftrightarrow$ $L(\mathcal{E})\neq\emptyset$.
\end{corollary}

We can solve the emptiness problem for $\mathcal{E}$ (see e.g. \cite{267878}), furthermore it is known that if $L(\mathcal{E})\neq\emptyset$,\linebreak then there exists a regular $(\Gamma_{\!\bot}\times St \times An)$-labeled $\Gamma$-tree $T_{reg}=(\Gamma^\ast, \lambda_{reg})$ and a deterministic finite automaton $\mathcal{A}_{reg}=(P,\Gamma, p_{in}, \delta_{reg}, f)$ where $f$ is an output function assigning to every state in $P$ a tuple from $\Gamma_{\!\bot}\times St \times An$, such that $T_{reg}$ is generated by $\mathcal{A}_{reg}$, i.e., the label of a node $\gamma\in \Gamma^\ast$ is  the output of the state $p$ reached after $\gamma$ has been processed by $\mathcal{A}_{reg}$, $\lambda_{reg}(\gamma)=f(\delta_{reg}^\ast(p_{in}, \gamma))$ where $\delta_{reg}^\ast$ is defined inductively as $\delta_{reg}^\ast(p,\varepsilon)=p$, $\delta_{reg}^\ast(p,wA)=\delta_{reg}(\delta_{reg}^\ast(p,w),A)$ for $w\in\Gamma^\ast$, $A\in\Gamma$.

Assume $L(\mathcal{E})\neq\emptyset$, then a winning pushdown strategy $\mathcal{S}$ for Player 0 in $\mathcal{G}(\mathcal{A})$ with the initial configuration $(q_{in}^\mathcal{A},\bot)$ can be derived from $\mathcal{A}_{reg}$ as follows. The states of $\mathcal{A}_{reg}$ are used as the pushdown alphabet of $\mathcal{S}$ with $p_{in}$ for the initial pushdown symbol, the states of $\mathcal{S}$ are the same as the states of $\mathcal{A}$. The pushdown strategy reads the letters chosen by Player 1 and outputs the next choice of Player 0 using the strategy encoded in the top of the stack. Formally, $\mathcal{S}=(Q^\mathcal{A},\Sigma, \Sigma, P, \delta^\mathcal{S}, q_{in}^\mathcal{A}, p_{in})$ where $\delta^\mathcal{S}$ contains the following transitions:

\begin{itemize}
 \item $q[p]\xrightarrow{a}q'[p]$, \hfill if $f(p)=(A,S,H)$ and $(q,a,N,q')\in S$
 \item $q[p]\xrightarrow{a}q'[p'][p]$, \hfill if $f(p)=(A,S,H)$, $(q,a,\downarrow_B,q')\in S$ and $\delta_{reg}(p,B)=p'$
 \item $q[p]\xrightarrow{a}q'\varepsilon$, \hfill if $f(p)=(A,S,H)$ and $(q,a,\uparrow,q')\in S$
\end{itemize}

\noindent Here we denote by $q[p]\xrightarrow{a}q'\pi$ the transition $\delta^\mathcal{S}(q,\varepsilon,p)=(q',\pi,a)$ if $q\in Q_0$, or $\delta^\mathcal{S}(q,a,p)=(q',\pi,\varepsilon)$ if $q\in Q_1$, i.e., if it is Player 0's turn, then an $\varepsilon$-transition is performed and the next choice $a\in\Sigma\cup\{\varepsilon\}$ for Player 0 is outputted and otherwise if $a$ is a letter chosen by Player 1, then it is processed with no output. 

Note that if Player 0 has no winning strategy in $\mathcal{G}(\mathcal{A})$, i.e., $L(\mathcal{E})=\emptyset$, then a winning pushdown strategy for Player 1 can be computed using this construction by swapping the roles of the players.

\section{Proof of Theorems}\label{Proof}
In the previous section we have described how a deterministic pushdown automaton realizing a winning strategy for the winner of a deterministic pushdown game can be constructed using the method from \cite{DBLP:conf/cav/KupfermanV00}. Now we explain how this technique can be adapted in order to solve parity games and stair parity games defined by DPDM, DVPM, realtime-DPDM and D1CM with winning strategies of corresponding types.

\subsubsection*{Stair Parity Games}

First we show how games with  stair conditions can be handled. For this, we construct an alternating two-way tree automaton which now evaluates the stair condition. 

\begin{definition}
 A \textit{stair A2TA} (StA2TA) $\mathcal{T}=(Q, \Sigma, q_{in}, \delta, col)$ over $\Sigma$-labeled $\Gamma$-trees has the same components as an A2TA. For a label $(q,\gamma)\in Q\times\Gamma^\ast$ and an infinite path $\rho\in(Q\times\Gamma^\ast)^\omega$ of a run $(T_r, \lambda_r)$ of $\mathcal{T}$ over some tree $T$ define $sh((q,\gamma))$ and $Steps_\rho$ accordingly. $\mathcal{T}$ accepts $T$ iff for all infinite paths $\rho$ of the run min$\{col(q) \ | \ q\in$ Inf$(\rho|_{Steps_\rho})\}$ is even, i.e., every infinite path of a run satisfies the stair parity condition.
\end{definition}

Let $\mathcal{A}$ be a parity-StDPDA defining a stair parity game $\mathcal{G}(\mathcal{A})$. We can assume $\mathcal{A}$ to be in normal form (every push-transition is of the form $\delta^\mathcal{A}(q,a,A)=(q',A'A)$). Define the StA2TA $\mathcal{B}$ from $\mathcal{A}$ along the lines of the previous section. It is required to test $\mathcal{B}$ for emptiness to determine the winner. 
To accomplish this the StA2TA $\mathcal{B}$ is transformed into an equivalent N1TA. This can be achieved by appropriate modifications of the automata $\mathcal{E}_i$ used in the construction of the previous section.

\paragraph*{From StA2TA to N1TA.}
The definitions of the strategy tree $\tau$ and the deterministic one-way tree automaton $\mathcal{E}_1$ over $(\Gamma_{\!\bot}\times St)$-labeled $\Gamma$-trees are not modified. 

For evaluating the stair condition almost all priorities seen during a finite detour are not relevant. The only important states of a detour, possibly constituting a position in $Steps_\rho$, are the first state, the state reached after the detour and all states seen at the same level as the first and the last state, i.e., at the same node in the tree. To keep track of these states the annotation is split into three components $\eta_1$, $\eta_2$ and $\eta_3$. For every node $\gamma\in\Gamma^\ast$, $\eta_1$ contains the information about all possible finite detours at $\gamma$,\linebreak $\eta_1\colon\Gamma^\ast\rightarrow\mathcal{P}(Q^\mathcal{B}\times Q^\mathcal{B})$, $(q,q')\in\eta_1(\gamma)$ means that there is a finite detour from $\gamma$ starting in state $q$ which comes back to $\gamma$ in state $q'$. The second component $\eta_2\colon\Gamma^\ast\rightarrow\mathcal{P}(Q^\mathcal{B}\times Q^\mathcal{B})$ contains the information about finite detours which return to $\gamma$ only one time, i.e., $(q,q')\in\eta_2(\gamma)$ means that there is a finite detour from $\gamma$ starting in state $q$ which comes back to $\gamma$ in state $q'$ and $\gamma$ was not visited during this detour elsewhere. The last component $\eta_3\colon\Gamma^\ast\rightarrow\mathcal{P}(Q^\mathcal{B} \times [k]\times Q^\mathcal{B})$ comprises for a finite detour the minimal priority of its $Steps$-positions, $(q,m,q')\in\eta_3(\gamma)$ means that there is a finite detour from $\gamma$ starting in $q$ and coming back to $\gamma$ in $q'$ with $m$ being the smallest priority seen on the $Steps$-positions of this detour.
More precisely, for a strategy tree $\tau$ a correct annotation $\eta=\eta_1\times\eta_2\times\eta_3$ has to satisfy the following conditions in every node $\gamma\in\Gamma^\ast$:

\begin{enumerate}
 \item if $(q,a,N,q')\in \tau(\gamma)$, then $(q,q')\in \eta_1(\gamma)$ and $(q,q')\in \eta_2(\gamma)$
 \item if $(q_1,q_2), (q_2,q_3)\in\eta_1(\gamma)$, then $(q_1,q_3)\in\eta_1(\gamma)$
 \item if $(q,a,\downarrow_{A},q_1)\in \tau(\gamma)$ and $(q_1,a',\uparrow,q')\in \tau(\gamma A)$, then $(q,q')\in \eta_1(\gamma)$ and $(q,q')\in \eta_2(\gamma)$
 \item if $(q,a,\downarrow_{A},q_1)\in \tau(\gamma)$ and $(q_1,q_2)\in\eta_1(\gamma A)$ and $(q_2,a',\uparrow,q')\in \tau(\gamma A)$, then $(q,q')\in \eta_1(\gamma)$ and $(q,q')\in \eta_2(\gamma)$
 \item if $(q,q')\in\eta_2(\gamma)$, then $(q,col^\mathcal{B}(q'),q')\in\eta_3(\gamma)$
 \item if $(q,m,q_1)\in\eta_3(\gamma)$, $(q_1,q')\in\eta_2(\gamma)$, then $(q,\text{min}\{m,col^\mathcal{B}(q')\},q')\in\eta_3(\gamma)$ 
\end{enumerate}

Let $An_1=An_2=\mathcal{P}(Q^\mathcal{B}\times Q^\mathcal{B})$ and $An_3=\mathcal{P}(Q^\mathcal{B}\times[k]\times Q^\mathcal{B})$. A deterministic one-way tree automaton $\mathcal{E}_2$ over $(\Gamma_{\!\bot}\times St \times An_1\times An_2\times An_3)$-labeled $\Gamma$-trees can be constructed which checks the correctness of the annotations $\eta_1$, $\eta_2$, $\eta_3$ for a strategy $\tau$.

Finally the automaton $\mathcal{E}_3$ evaluating the parity condition has to be modified properly, so that it reads the strategy and the third component of the annotation $(St\times An_3)$. Note that due to the nature of $\eta_3$ containing the essential information about the $Steps$-positions of finite detours, in fact the stair parity condition gets evaluated, therefore ensuring the correctness of the reduction. 

\begin{corollary}
 For every StA2TA $\mathcal{B}$ there exists an equivalent N1TA $\mathcal{E}'$ such that $L(\mathcal{B})=L(\mathcal{E}')$.
\end{corollary}

\subsubsection*{Visibly Games}

Let $\mathcal{A}$ be a parity-DVPA or a parity-StDVPA defining a visibly pushdown game $\mathcal{G}(\mathcal{A})$ and let $\mathcal{S}$ be a winning pushdown strategy w.l.o.g. for Player 0 in $\mathcal{G}(\mathcal{A})$ constructed as above. Note that the stack height of $\mathcal{S}$ is controlled by the input and output letters, i.e., $\mathcal{S}$ performs a push-transition if the processed symbol is $a\in\Sigma_c$, a pop-transition is performed if $a\in\Sigma_r$ and a skip-transition is performed if $a\in\Sigma_{int}$. However, $\mathcal{S}$ is not a VPA yet, since for VPA the use of the stack is even more restricted (there is no access to the top of the stack on processing calls and internal actions). Nevertheless, $\mathcal{S}$ can easily be converted into a VPA $\mathcal{S}_{vis}$ defining a winning visibly pushdown strategy for Player 0 by extending the stack alphabet and the set of states, $\Gamma_{vis}= (P\times P)\cup\{p_{in}\}$ and $Q_{vis}=Q^\mathcal{A}\times P$. 

\subsubsection*{Realtime Games}

The problem that arises with realtime games is that the transformation into the normal form introduces $\varepsilon$-transitions. This can be resolved at the stage when the winning strategy is derived from the finite automaton generating the regular tree.

Let $\mathcal{A}$ be a realtime parity-DPDA or a realtime parity-StDPDA defining a realtime pushdown game $\mathcal{G}(\mathcal{A})$. First $\mathcal{A}$ is translated into an equivalent automaton $\mathcal{A}_{norm}$ in normal form using the usual construction, $Q_{norm}=Q^{\mathcal{A}}\cup(Q^{\mathcal{A}}\times \Gamma^m)$ where $m=\text{max}\{|\gamma| \ | \ (q', \gamma)\in\delta^{\mathcal{A}}(q, a, A) \}$ and for $A\in\Gamma_{\!\bot}$, $A_1,...,A_n\in \Gamma$, $q,q'\in Q^{\mathcal{A}}$ and $a\in\Sigma$:

\begin{itemize}
 \item $\delta_{norm}(q,a,A) = (\langle q',A_1...A_n\rangle, \varepsilon)$, \hfill if $\delta^{\mathcal{A}}(q,a,A) = (q',A_1...A_n)$, $A\neq\bot$
 \item $\delta_{norm}(q,a,\bot) = (\langle q',A_1...A_n\rangle, \bot)$, \hfill if $\delta^{\mathcal{A}}(q,a,\bot) = (q',A_1...A_n\bot)$
 \item $\delta_{norm}(\langle q,A_1...A_n\rangle,\varepsilon,A) = (\langle q,A_1...A_{n-1}\rangle, A_nA)$, \hfill if $n>0$
 \item $\delta_{norm}(\langle q,\varepsilon\rangle,\varepsilon,A) = (q, A)$
\end{itemize}

The A2TA (StA2TA) $\mathcal{B}$ is constructed from $\mathcal{A}_{norm}$ and checked for emptiness. We deduce from the finite automaton $\mathcal{A}_{reg}$ generating the regular tree $T_{reg}=(\Gamma^\ast, \lambda_{reg})$ a realtime winning strategy $\mathcal{S}$. The idea is to merge subsequent $\varepsilon$-transitions with the foregoing non-$\varepsilon$-transition to one non-$\varepsilon$-transition. For this, it is necessary to have access to the two topmost stack symbols, thus the stack alphabet is extended to $(P\times P) \cup\{p_{in}\}$. For a state $q$ and the topmost stack symbol $(p,\overline{p})$ with $f(p)=(A,S,H)$ and $(q,a,\uparrow,\langle q',A_1...A_n\rangle)\in S$, $\delta^\mathcal{S}$ contains the following transition:

\begin{flushleft}
$q(p,\overline{p})\xrightarrow{a}q'(p_1,p_2)(p_2,p_3)...(p_{n-1},p_n)(p_n,\overline{p})$, \hfill if $\delta_{reg}(\overline{p},A_n)=p_n$ and $\delta_{reg}(p_i,A_{i-1})=p_{i-1}$,\\
\raggedleft{for $i=n, n-1,...,2$}                                                                                                                                                                                                    \end{flushleft}

\noindent For the other cases the transitions are defined similarly. By this means, $\mathcal{S}$ contains no $\varepsilon$-transitions, since the artificial $\varepsilon$-transitions generated during the transformation of $\mathcal{A}$ into normal form are eliminated. Hence, $\mathcal{S}$ is realtime.

\subsubsection*{One-Counter Games}

Given a parity-D1CA (parity-StD1CA) $\mathcal{A}=(Q^\mathcal{A}, \Sigma, \{A\}, \delta^\mathcal{A}, q_{in}^\mathcal{A}, \bot, col^\mathcal{A})$  in normal form defining a one-counter game $\mathcal{G}(\mathcal{A})$, construct A2TA (StA2TA) $\mathcal{B}$ as in previous section. Note that due to the fact that the pushdown alphabet is a singleton, in this special case $\mathcal{B}$ can be viewed as an alternating two-way \textit{word} automaton. Player 0 has a winning strategy in $\mathcal{G}(\mathcal{A})$ from the initial configuration if and only if $\bot A^\omega \in L(\mathcal{B})$. This is checked as in previous cases by a reduction to a one-way automaton $\mathcal{E}$, now a one-way \textit{word} automaton.

Assume $L(\mathcal{E})\neq\emptyset$, then there exists an ultimately periodic word $w\in(\Gamma_{\!\bot}\times St\times An)^\omega$ and a deterministic finite automaton $\mathcal{A}_{reg}=(P,p_{in},\delta_{reg},f)$ where $P=\{p_1, ..., p_n\}$ with $p_{in}=p_1$, $\delta_{reg}(p_i)=p_{i+1}$ for all $1\leq i < n$ and $\delta_{reg}(p_n)=p_l$ for one $l\in\{1,...,n\}$ and $f\colon P \rightarrow (\Gamma_{\!\bot}\times St\times An)$, such that $w$ is generated by $\mathcal{A}_{reg}$, i.e., $w(i)=f(\delta_{reg}^i(p_{in}))$.

From $\mathcal{A}_{reg}$ we derive a winning one-counter strategy $\mathcal{S}$. Note that employing $P$ as the stack alphabet of $\mathcal{S}$ as in previous cases does not lead to a counter strategy, since there is only one stack symbol available. Instead of that, we additionally store the information encoded in $P$ into the states of $\mathcal{S}$ and use the stack in order to count the number of times $\mathcal{A}_{reg}$ goes into its loop. Let $f(p_i)=(A_i, S_i, H_i)$ for $1 \leq i \leq n$. Formally, $\mathcal{S}=(Q^\mathcal{A}\times P,\Sigma, \Sigma, \{A\}, \delta^\mathcal{S}, (q_{in}^\mathcal{A},p_{in}), \bot)$ and $\delta^\mathcal{S}$ contains the following transitions where $X\in\{A,\bot\}$:

\begin{itemize}
 \item $(q,p_i)X\xrightarrow{a}(q',p_i)X$, \hfill if $(q,a,N,q')\in S_i$

 \item $(q,p_i)X\xrightarrow{a}(q',p_{i+1})X$, \hfill if $(q,a,\downarrow,q')\in S_i$ and $1\leq i < n$
 \item $(q,p_n)X\xrightarrow{a}(q',p_{l})AX$, \hfill if $(q,a,\downarrow,q')\in S_n$
 
 \item $(q,p_i)X\xrightarrow{a}(q',p_{i-1})X$, \hfill if $i\neq 1$, $i\neq l$, and $(q,a,\uparrow,q')\in S_i$
 \item $(q,p_l)\bot\xrightarrow{a}(q',p_{l-1})\bot$ and $(q,p_l)A\xrightarrow{a}(q',p_n)\varepsilon$, \hfill if $(q,a,\uparrow,q')\in S_l$
\end{itemize}

\noindent In the case of a push-transition the second component $p_i$ is properly updated to $\delta_{reg}(p_i)$ and in the case of a pop-transitions it should be updated to $\delta_{reg}^{-1}(p_i)$. The crucial point is the state $p_l$ where the loop of $\mathcal{A}_{reg}$ is completed, since it has two predecessor states,  $\delta_{reg}(p_{l-1})=\delta_{reg}(p_{n})=p_l$. Thus, for performing a pop-transition it is required to know, if to return to $p_{l-1}$ or to $p_n$. The appropriate update is obtained by observing the stack which is increased every time $\mathcal{S}$ proceeds from $p_n$ to $p_l$. If the stack is empty, then $p_l$ was reached from $p_{l-1}$, otherwise if the stack is not empty, then the loop was completed by the transition from $p_n$ to $p_l$, in this case $\mathcal{S}$ returns to $p_n$ and decreases the stack size, if a pop-transition is proposed to be performed by the strategy $S_l$.
 
\subsubsection*{Blind One-Counter and Visibly One-Counter Games}

Consider a blind one-counter game defined by the following parity-DB1CA $\mathcal{A}=(\{q_0,q_1,q_2,q_3,q_4\}, \linebreak \{a,b,c,d\}, \{A\}, \delta, q_{0}, \bot, col)$ with $Q_0=\{q_2,q_3\}$ and $Q_1=\{q_0,q_1,q_4\}$, $\delta(q_0,a,X)=(q_0,AX)$, \linebreak $\delta(q_0,b,A)=\delta(q_1,b,A)=(q_1,\varepsilon)$, $\delta(q_1,c,X)=(q_2,X)$, $\delta(q_2,a,X)=(q_3,X)$, $\delta(q_2,b,X)=(q_4,X)$, $\delta(q_3,c,X)=(q_4,AX)$, $\delta(q_4,c,X)=(q_3,AX)$, $\delta(q_3,d,A)=(q_3,A)$, $\delta(q_4,d,A)=(q_4,A)$, for $X\in\{A,\bot\}$ and $col(q_0)=col(q_1)=2$, $col(q_2)=col(q_3)=0$, $col(q_4)=1$.

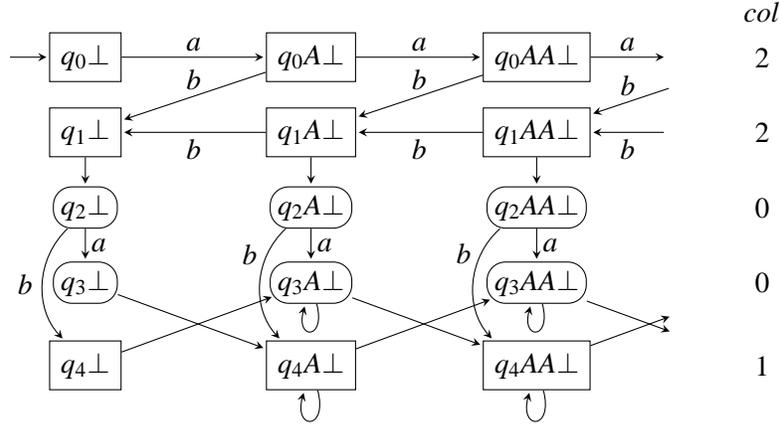
\begin{figure}
\begin{center}
\begin{tikzpicture}[shorten >=1pt,node distance=3cm,auto,  initial text=, >=stealth]

\node[state, text=gray, draw=none] at (0,.6)	(0p1) {};
\node[state, text=gray, draw=none] 		(0p2) [right of=0p1] {};
\node[state, text=gray, draw=none] 		(0p3) [right of=0p2] {};

\tikzstyle{every state}=[fill=none,minimum size=0.2mm ,draw,inner sep=1.5mm,rectangle]

\node[state,initial,draw] at (0,0) 	(1p1) {$q_0\bot$};
\node[state,  draw] 		(1p2) [right of=1p1] {$q_0A\bot$};
\node[state,  draw] 		(1p3) [right of=1p2] {$q_0AA\bot$};

\node[state,  draw] at (0,-1)	(2p1)  {$q_1\bot$};
\node[state,  draw] 		(2p2) [right of=2p1] {$q_1A\bot$};
\node[state,  draw] 		(2p3) [right of=2p2] {$q_1AA\bot$};

\tikzstyle{every state}=[fill=none,minimum size=0.2mm ,draw,inner sep=1mm,rectangle, rounded corners=7pt]

\node[state,  draw] at (0,-2)	(3p1)  {$q_2\bot$};
\node[state,  draw] 		(3p2) [right of=3p1] {$q_2A\bot$};
\node[state,  draw] 		(3p3) [right of=3p2] {$q_2AA\bot$};

\node[state,  draw] at (0,-3)	(4p1)  {$q_3\bot$};
\node[state,  draw] 		(4p2) [right of=4p1] {$q_3A\bot$};
\node[state,  draw] 		(4p3) [right of=4p2] {$q_3AA\bot$};

\tikzstyle{every state}=[fill=none,minimum size=0.2mm ,draw,inner sep=1.5mm,rectangle]

\node[state,  draw] at (0,-4.1)	(6p1)  {$q_4\bot$};
\node[state,  draw] 		(6p2) [right of=6p1] {$q_4A\bot$};
\node[state,  draw] 		(6p3) [right of=6p2] {$q_4AA\bot$};

\tikzstyle{every state}=[fill=none,minimum size=0.2mm ,draw,inner sep=1mm,circle]

\node[state, draw=none] (6p2c) [right of=6p1] {};
\node[state, draw=none] (6p3c) [right of=6p2] {};
\node[state, draw=none] (6p4c) [right of=6p3] {};

\node[state,  draw=none] 		(0p) [right of=0p3] {$col$};
\node[state,  draw=none] 		(1p) [right of=1p3] {2};
\node[state,  draw=none] 		(2p) [right of=2p3] {2};
\node[state,  draw=none] 		(3p) [right of=3p3] {0};
\node[state,  draw=none] 		(4p) [right of=4p3] {0};
\node[state,  draw=none] 		(6p) [right of=6p3] {1};

\path[->]
(1p1) edge node[above=-.5mm]{$a$} (1p2)
(1p2) edge node[above=-.5mm]{$a$} (1p3)
(1p3) edge[shorten >=1cm] node[above=-.5mm, near start]{$a$} (1p)
;

\path[->]
(2p2) edge node[below=-.5mm]{$b$} (2p1)
(2p3) edge node[below=-.5mm]{$b$} (2p2)
(2p) edge[shorten <=1cm] node[below=-.5mm, near end]{$b$} (2p3)
;

\path[->]
(1p2) edge node[above=-.5mm]{$b$} (2p1)
(1p3) edge node[above=-.5mm]{$b$} (2p2)
(1p) edge[shorten <=1cm] node[above=-.5mm, near end]{$b$} (2p3)
;

\path[->]
(2p1) edge node[right=-.5mm]{} (3p1)
(2p2) edge node[right=-.5mm]{} (3p2)
(2p3) edge node[right=-.5mm]{} (3p3)
;

\path[->]
(3p1) edge node[right=-.5mm]{$a$} (4p1)
(3p2) edge node[right=-.5mm]{$a$} (4p2)
(3p3) edge node[right=-.5mm]{$a$} (4p3)
;

\path[->]
(3p1.230) edge[bend right=45] node[left]{$b$} (6p1.130)
(3p2.220) edge[bend right=45] node[left, near start]{$b$} (6p2.140)
(3p3.210) edge[bend right=45] node[left, near start]{$b$} (6p3.150)
;

\path[->]
(4p1) edge node{} (6p2)
(4p2) edge node{} (6p3)
(4p3) edge[shorten >=1cm] node{} (6p)
;

\path[->]
(6p1) edge[shorten >=-0.03cm] node{} (4p2)
(6p2) edge[shorten >=-0.05cm] node{} (4p3)
(6p3) edge[shorten >=1cm] node{} (4p)
;

\path[->]
(4p2) edge[loop below] ()
(4p3) edge[loop below] ()
;

\path[->]
(6p2) edge[loop below] ()
(6p3) edge[loop below] ()
;
\end{tikzpicture}
\end{center}
    \caption{Blind one-counter game.}
    \label{BlindGame}
\end{figure}

The game graph $G(\mathcal{A})$ is depicted in Figure \ref{BlindGame} where for better readability the labels of $c$- and $d$-transitions are omitted.
Configurations of Player 1 are indicated by rectangles and Player 0 nodes are rounded. Note that every transition which is enabled with empty stack is also enabled with nonempty stack. Player 1 begins by building up a finite prefix $a^nb^mc$ with $m\leq n$. He should not take an infinite number of $a$'s and stay in the initial state $q_0$ forever, since $col(q_0)=2$. After a prefix $a^nb^mc$ is provided, Player 0 has to decide whether to pick $a$ or $b$. Player 0 will win if he can force reaching a loop in state $q_3$, on the other hand he will loose if a loop in $q_4$ is reached where Player 1 can stay forever by choosing $d$. Hence, a winning strategy for Player 0 is to pick $a$ being in state $q_2$ if the prefix constructed by Player 1 contains more $a$'s than $b$'s and to pick $b$ if it contains equal number of $a$'s and $b$'s. This strategy can easily be realized by a DPDA.

We use a simple language theoretic argument to show that there exists no DB1CA $\mathcal{S}$ realizing a winning strategy for Player 0 in this game.

\begin{lemma}
 The language $L=\{a^nb^nc \ | \ n>0\}$ is not accepted by any DB1CA.
\end{lemma}

\noindent Since every winning strategy for Player 0 has to decide whether the prefix chosen by Player 1 is contained in $\{a^nb^nc \ | \ n>0\}$ or in $\{a^nb^mc \ | \ m<n\}$, with the above lemma it is clear, that this cannot be realized by any DB1CA.

With a similar argument it can be shown that visibly one-counter strategies do not suffice to solve visibly one-counter games. Consider the following DV1CL$_\omega$-game which we describe informally without giving the detailed definitions of a visibly one-counter inducing this game. Player 1 begins by constructing a prefix $c^na$ with $n\geq2$. Then Player 0 responds by a sequence $r^mar^2$ followed by $a^\omega$. Player 0 wins a play if $m=n-2$, 
thus, the winning condition is given by $L_{win}=\{c^nar^{n-2}ar^2a^\omega \ | \ n\geq2\}\subseteq\{c,r,a\}^\omega$. A winning visibly pushdown strategy for Player 0 can be constructed as for any deterministic visibly pushdown game. However, there exists no visibly one-counter implementing a winning strategy for \linebreak Player 0.

\begin{lemma}
 The language $L=\{c^nar^{n-2} \ | \ n\geq2\}$ is not accepted by any DV1CA.
\end{lemma}

\noindent Since every winning strategy for Player 0 has to find the correct position to place the $a$, with the above lemma it is clear that this cannot be done by any DV1CA.

\section{Conclusion}\label{Conclusion}

We exhibited several types of pushdown games which by a uniform proof method turned out to be solvable by pushdown strategies of corresponding format, namely parity games as well as stair parity games played on game arenas defined by deterministic, deterministic visibly, deterministic realtime pushdown machines and deterministic one-counter machines. Furthermore, two types of pushdown games were indicated where strategies of corresponding format emerged to be not sufficient, namely the blind one-counter and the visibly one-counter games. 

This result raises the question concerning the abstract reasons for the transfer from game specifications to solutions of the same format. Can we precisely separate the classes of pushdown games where solvability with winning strategies of corresponding format is guaranteed from those classes where this is not the case? We add some remarks on a result developed in detail in a future paper.

Let $\mathcal{F}$ be a \textit{format} of a PDM $\mathcal{M}$. We call a parity game (stair-parity game) \textit{$\mathcal{F}$-definable} if there exists a parity-DPDA (parity-StDPDA) of format $\mathcal{F}$ inducing the game. We call a pushdown strategy \textit{$\mathcal{F}$-definable} if it can be realized by a DPDA of format $\mathcal{F}$.

We give a sufficient condition for solvability of pushdown games with winning strategies of corresponding format. Based on the observation that the essential task of a pushdown strategy is to navigate on a regular infinite tree 
we introduce the notions of \textit{$\mathcal{F}$-guidability} and \textit{adequacy} of a format $\mathcal{F}$ and obtain that for every adequate format $\mathcal{F}$, $\mathcal{F}$-definable parity games and $\mathcal{F}$-definable stair parity games are determined with $\mathcal{F}$-definable winning strategies.

\paragraph{Acknowledgements.}
This work was initiated in the author's diploma thesis under the supervision of Wolfgang Thomas. I wish to thank him for his advice and suggestions. Also, I want to thank the anonymous referees for their remarks.

\bibliographystyle{eptcs}
\bibliography{LitVerz}

\end{document}